\documentclass[iop]{emulateapj}
\begin{document}

\newcommand{\es}{\ensuremath{\hspace{2pt} \rm ergs \hspace{2pt} \rm s^{-1}}}

\title{Supernova Light Curves Powered by Fallback Accretion}
\shorttitle{Accretion Powered Supernova Light Curves}
\shortauthors{Dexter \& Kasen}

\author{Jason Dexter}
\affil{Departments of Physics and Astronomy,
University of California, Berkeley, CA 94720, USA}
\email{jdexter@berkeley.edu}

\author{Daniel Kasen}
\affil{Departments of Physics and Astronomy, University of California, Berkeley, CA, USA}
\affil{Nuclear Science Division, Lawrence Berkeley National Laboratory, 1 Cyclotron Road, Berkeley, CA, USA}

\begin{abstract}
Some fraction of the material ejected in a core collapse supernova explosion may remain bound to the compact remnant, and eventually turn around and fall back.   We show that the late time ($\ga $ days) power associated with the accretion of this ``fallback'' material may significantly affect the  optical  light curve, in some cases producing super-luminous or otherwise peculiar supernovae. We use spherically symmetric hydrodynamical models to estimate the accretion rate at late times for a range of progenitor masses and radii and explosion energies. The accretion rate onto the proto-neutron star or black hole decreases as $\dot{M} \propto t^{-5/3}$ at late times, but its normalization can be significantly enhanced at low explosion energies, in very  massive stars, or if a strong reverse shock wave forms at the helium/hydrogen interface in the progenitor. If the resulting super-Eddington accretion drives an outflow which thermalizes in the outgoing ejecta,  the supernova debris will be re-energized at a time when photons can diffuse out efficiently.  The resulting light curves are different and more diverse than previous fallback supernova models which ignored the input of accretion power and produced short-lived, dim transients.  The possible outcomes when fallback accretion power is significant include super-luminous ($\gtrsim 10^{44} \es$) Type II events of both short and long durations, as well as luminous Type I events from compact stars that may have experienced significant mass loss. Accretion power may unbind the remaining infalling material, causing a sudden decrease in the brightness of some long duration Type II events. This scenario may be relevant for explaining some of the recently discovered classes of peculiar and rare supernovae. 
\end{abstract}

\maketitle

\section{Introduction}

Ongoing optical  surveys have discovered new classes of supernovae (SNe), including sub-luminous \citep[e.g.,][]{Perets:2009p3705,foleyetal2009,kasliwaletal2010} and super-luminous \citep[e.g.,][]{quimbyetal2011,galyam2012} events. Many of these events are difficult to explain in the context of standard models for which radioactive decay powers the optical light curve \citep[e.g., SN 2008es,][]{gezarietal2009}. Alternative explanations fall into two main categories.  Overluminous Type~IIn SNe with narrow hydrogen absorption lines are thought to be powered, at least in part, by the interaction of the supernova ejecta with circumstellar material, effectively re-thermalizing the supernova shock energy \citep{smithmccray2007,chevalierirwin2011, moriya2012,galyam2012}.  Alternatively,  the SN ejecta may be re-energized by the spindown power of a rapidly rotating magnetar which formed in the core collapse  \citep[][hereafter KB10]{Kasen:2010p7056}.   For either of these two mechanisms to significantly modify the SN light curve, the energy input must occur at relatively late times (weeks to months) when radiative diffusion through the ejecta is efficient.

Accretion onto a central compact remnant represents another potential means of injecting large amounts of energy in either successful or failed supernova explosions. Compact object accretion is associated with large-scale outflows in neutron stars \citep{fenderetal2004} and stellar mass black holes \citep[microquasars, e.g.,][]{mirabelrodriguez1998}, and these outflows can carry away as much as $\sim 10\%$ of the gravitational binding energy of the infalling material. This is particularly true when the accretion flow is rotationally supported and radiatively inefficient \citep{narayanyi1994,blandfordbegelman1999}. 

In ``failed" SNe, the entire star accretes onto the central remnant, a black hole. If the progenitor lacks sufficient angular momentum to form a disk and hence tap the available accretion energy, these events are ``unnovae'' \citep{Kochanek:2008p20323}, i.e., stars disappearing suddenly from view.   In the opposite case where even the iron core becomes rotationally supported, the accretion energy may power a long-duration gamma ray burst (GRB)  \citep[the collapsar mechanism,][]{Woosley:1993p18359,MacFadyen:1999p9378}. Much longer gamma ray transients may also be possible if either the mantle \citep{MacFadyen:2001p9343} or the hydrogen envelope \citep[][hereafter QK12]{Woosley:2011p19913,Quataert:2012p20328} becomes rotationally supported and drives a relativistic jet. The timescale associated with the energy injection corresponds roughly to the free-fall time of a stellar layer, about $\sim 0.1$~s for the iron core, but as long as $\sim 1$ yr for the hydrogen envelope of a red giant. Powerful winds from the accretion disk  may eventually provide sufficient energy to turn the failed SN into a successful one, exploding the remainder of the star   \citep{Milosavljevic:2010p18021,Lindner:2011p20292}.

In successful SNe,  accretion  from the ``fallback'' of the fraction of material remaing bound can be significant as well. The fallback may influence the resulting nucleosynthesis \citep{Colgate:1971p19161} or delay the pulsar mechanism in a young proto-neutron star such as in SN 1987A \citep{Michel:1988p19164}. Early time fallback may also provide a link between the explosion mechanism and the remnant mass distribution \citep{fryeretal2012} or alter the radiated neutrino spectrum \citep{Fryer:2009p17660}. For red supergiant (RSG) progenitors with typical explosion energies ($\simeq 10^{51} \rm erg$), the fallback mass is relatively small ($\sim 0.1~M_\odot$).  However, in more compact stars (e.g., blue supergiants, BSGs) the formation of a strong reverse shock at the H/He interface can decelerate the ejecta and enhance the fallback mass to $\sim 2 M_\odot$ \citep{Chevalier:1989p9522,Zhang:2008p9515}.   For weak explosions, most of the star may fall back, with only a small fraction of the mass ejected in a dim SN \citep{Moriya:2010p19157}.

While the dynamics of supernova fallback have been studied in a various contexts, little has been said about how fallback may impact the optical SN light curve.  The energy released from fallback accretion may profoundly affect what we observe, if two conditions are met.  First, the accretion energy must be injected at relatively late times ($ \ga$ days) otherwise it will be largely degraded by adiabatic expansion.  Such late time accretion may be possible for progenitors with extended envelopes, or for those where a reverse shock develops and gradually slows the inner layers of ejecta.  Second, the accretion energy must be thermalized within the SN ejecta.  This is likely to occur if the energy injection takes the form of a nearly isotropic disk wind.  If, on the other hand, the energy is in a beamed relativistic jet, we must consider whether the jet can breakout of the ejecta (perhaps producing a GRB) or whether it is trapped and thermalized in the interior.  When these two conditions are met,  fallback should produce a peculiar optical light curve, powered directly by the accretion energy.

We study the impact of late time fallback accretion of SN light curves, and suggest that the wide range of potential  events -- from sub luminous to super-luminous --  may be of relevance in explaining recent observations of peculiar SNe. In \S\ref{accenergy}, we crudely estimate the efficiency of fallback-accretion-driven outflows. We numerically calculate accretion rates for a wide range of stellar progenitors to explore the variety of outcomes for accretion powered supernova light curves (\S\ref{outcomes}), including sample light curves and their comparison to some recent unusual events  (\S\ref{sec:candidate-events}). We also attempt to address the various requirements for these events to occur in Nature: the interaction of the outflows with both the infalling material and the outgoing ejecta (\S\ref{interactejecta}), and angular momentum and disk formation (\S\ref{angmom}). The major results are summarized in \S\ref{summary}.

\section{Accretion Energy}
\label{accenergy}

At both low and high accretion rates compared to $L_{\rm edd} / c^2$, where $L_{\rm edd}$ is the Eddington luminosity, accretion flows onto compact objects become hot and geometrically thick due to their inability to cool efficiently ($t_{\rm cool} > t_{\rm infall}$). Such radiatively inefficient accretion flows are expected to produce large-scale outflows \citep{narayanyi1994,blandfordbegelman1999,igumenshchevabramowicz2000,penetal2003,begelman2012,mckinneyetal2012} and/or Poynting flux dominated jets \citep{devilliersetal2005,mckinney2006}. This behavior is observed in the accretion flow onto the Galactic center black hole, Sagittarius A*, where the accretion rate at the Bondi radius \citep[e.g.,][]{baganoffetal2003,quataert2004} is several orders of magnitude larger than that onto the black hole \citep[e.g.,][]{marroneetal2007}. The fallback accretion rate following a successful supernova explosion is highly super-Eddington and extremely optically thick to photons. For all timescales of interest here ($\gtrsim 1000$s after the explosion), the disk is not dense enough to cool by neutrino emission \citep{Kohri:2005p19177}. We expect then that it should be radiatively inefficient, geometrically thick, and should drive large-scale outflows. 

The resulting mass outflow rate can be estimated following \citet{Kohri:2005p19177} by assuming that the accretion rate increases as some power of radius
\begin{equation}
\dot{M} (r) = \dot{M}_{\rm fb} \left(\frac{r}{r_{\rm fb}}\right)^s,
\end{equation}

\noindent where $\dot{M}_{\rm fb}$ and $r_{\rm fb}$ are the accretion rate and radius at the outer disk edge, and  $0 < s < 1$.   We will write the radius in units  of the
Schwarzchild radius, $r = R c^2/G M$.
The outflow speed should be comparable to the escape speed, $v_w \sim c/\sqrt{2 r}$, and the energy in each disk annulus is,

\begin{equation}
d\dot{E}_w = \xi d\dot{M} \frac{c^2}{2 r} = \frac{\xi s}{2} \frac{\dot{M}_{\rm fb} c^2}{r_{\rm fb}^s r^{2-s}} dr,
\end{equation}

\noindent where $\xi$ parameterizes our ignorance of the outflow physics, such as the fraction of fallback mass that is blown out again. The actual value of $s$ is highly uncertain, but $s=1$ is a reasonable choice \citep[e.g.,][]{igumenshchevabramowicz2000,penetal2003,begelman2012,mckinneyetal2012}, in which case the total outflow rate integrated over disk radius is,

\begin{equation}
\dot{E}_w = \frac{\xi \dot{M}_{\rm fb} c^2}{2 r_{\rm fb}} \log \left(\frac{r_{\rm fb}}{r_{\rm in}}\right),
\end{equation}

\noindent where $r_{\rm in}$ is the inner disk edge, either the black hole event horizon or the surface of the proto neutron star. For typical parameters we take $r_{\rm fb}=100$ ($\sim 10^8 \rm cm$) and $\xi=0.1$. These choices give an outflow energy $\dot{E}_w = \epsilon \dot{M}_{\rm fb} c^2$ with $\epsilon \sim 10^{-3}$. For the parameters considered here, other choices of $s > 0$ give similar results. If the outflow is instead a jet launched from near the inner disk edge, the outflow energy is $\dot{E}_{j} = \beta \dot{M}_{\rm bh} c^2$, where a conventional choice  is $\beta \sim 0.1$, although depending on the accreted magnetic field geometry this value could be much larger \citep{mckinneyetal2012}. Since $\dot{M}_{\rm bh} = \dot{M}_{\rm fb} (r_{\rm fb} / r_{\rm in})$, the resulting energy injection, $\epsilon \dot{M}_{\rm fb} c^2$ would be nearly identical to the case of a disk wind. This is the scenario discussed for failed supernova explosions by \citet{Woosley:2011p19913} and QK12. The results presented below only depend on the energy injection rate and thus are the same for either a disk or a jet. The results are also expected to be insensitive to whether the central object is a proto neutron star or black hole. In \S \ref{interactejecta} we discuss the dissipation of accretion energy in the infalling material and outgoing ejecta, and its implications for the viability of wind and jet scenarios.

We use $\epsilon=10^{-3}$ throughout, although we discuss disk formation and size in \S \ref{angmom}. The outflow energy, either from a wind or jet, is then set by the fallback accretion rate. 

\begin{figure}
\epsscale{1.0}
\plotone{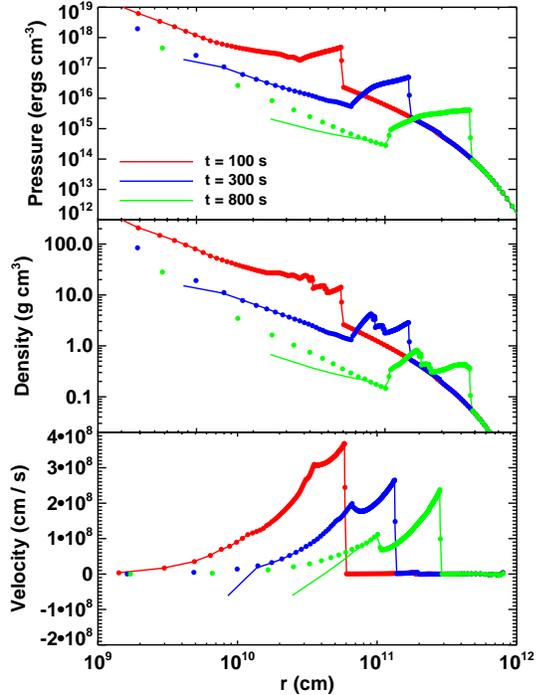}
\caption{\label{pvd}Pressure (top), density (middle), and velocity (bottom) vs. radius at $t = 100$s, $300$s, and $800$s for an $E_0 = 1.2 \times 10^{51}$ erg explosion of a $25 M_\sun$ zero metallicity progenitor using the fallback (lines) and piston (points) inner boundary conditions. The two methods are in good agreement away from the inner boundary, and the results are similar to \citet{Zhang:2008p9515} Figure 1. The points have been down-sampled by a factor of 5 for clarity.}
\end{figure}

\begin{figure}
\epsscale{1.0}
\plotone{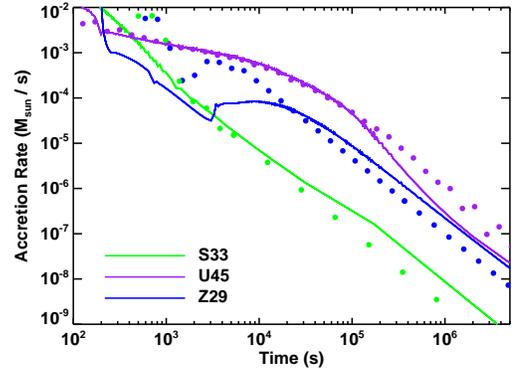}
\caption{\label{mdot}Numerical (lines) and semi-analytic (points) fallback accretion rates for three models from Table \ref{tab:events}. For the small progenitor in S33, the asymptotic scaling $\dot{M} \propto t^{-5/3}$ applies after the first $\simeq 100$s. At much lower explosion energies (U45), the entire accretion rate curve is well described by freefall. In BSG progenitors (Z29), the accretion rate can be significantly enhanced at late times by material re-captured by the reverse shock formed at the H/He interface.}
\end{figure}

\begin{figure*}
\epsscale{0.8}
\plotone{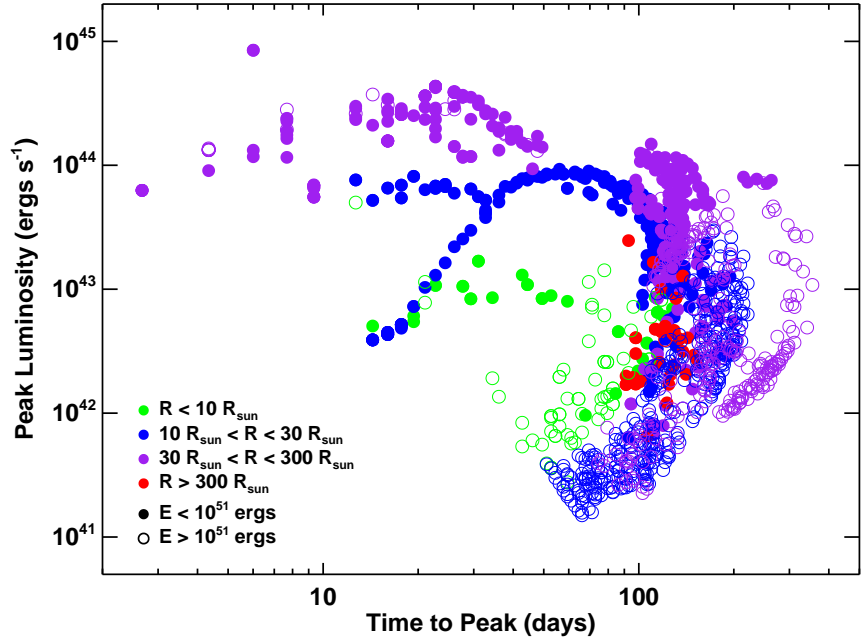}
\caption{\label{lptpl}Peak luminosity vs. time to peak for all events, measured from light curves calculated with the methods in Appendix A. The points are only shown when the peak luminosity is larger than the standard thermal supernova luminosity (Eq. ~\ref{eq:3}). Luminous, long duration Type II events come from high energy explosions in massive BSGs, some with strong reverse shocks. Superluminous events with durations $\simeq 2-40$ days come from weak explosions with low ejecta masses. Luminous Type I events are from weak explosions in compact initial stars.}
\end{figure*}

\begin{figure}
\epsscale{1.0}
\plotone{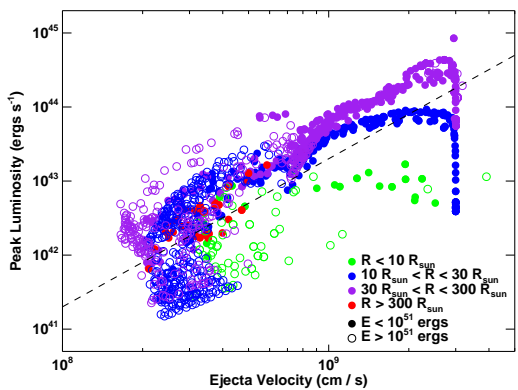}
\caption{\label{tpvf}Peak luminosity vs. $v_f$ for the events in Figure \ref{lptpl}. The peak luminosity scales roughly with final ejecta velocity as $L_p \propto v_f^2$.}
\end{figure}

\begin{table*}
\caption{Sample Event Parameters \label{tab:events}}
\begin{small}
\begin{center}
\begin{tabular}{lccccccccc}
        \tableline
	\tableline
Name & $M_{\rm ZAMS}$ ($M_\sun$) & $M_{\rm SN}$ ($M_\sun$) & $Z$ ($Z_\sun$) & $E_{\rm exp}$ ($10^{51} \rm ergs$) & $M_{\rm ej}$ ($M_\sun$) & $M_{\rm rem}$ ($M_\sun$) & $v_{\rm ej}$ (km / s) & $v_f$ (km / s) & Comparison\\
        \tableline
S33 & $33.0$ & $11.4$ & $1$ & $0.34$ & $2.1$ & $9.2$ & $2800$ & $24000$ & SN 1998bw \\
S39 & $39.0$ & $8.49$ & $1$ & $0.21$ & $1.1$ & $7.4$ & $3000$ & $17000$ & SN 2008D \\
S39O & $39.0$ & $8.49$ & $1$ & $0.055$ & $0.39$ & $8.1$ & $2600$ & $11000$ & SN 2010X \\
U45 & $45$ & $44.7$ & $10^{-4}$ & $1.0 \times 10^{-3}$ & $0.45$ & $44$ & $340$ & $13000$ & SN 2008es\\
U60 & $60$ & $59.2$ & $10^{-4}$ & $1.0$ & $43$ & $17$ & $1100$ & $5500 $ & --- \\
Z29 & $29$ & $28.8$ & $0$ & $1.1$ & $22$ & $6.7$ & $1600$ & $4700$ & --- \\
	\tableline
\end{tabular}
\end{center}
\end{small}
\end{table*}

\section{Fallback Accretion}
\label{sec:numer-calc}

\subsection{Numerical Hydrodynamics} 
We estimate the fallback accretion rate by simulating supernova explosions using a 1D lagrangian finite difference hydrodynamics code. The code uses a  staggered mesh and artificial viscosity shock prescription \citep{castor2004}. The artificial viscosity parameter is chosen to smooth shocks over $\simeq 7$ zones. This is fairly diffusive, helping with code stability but gives nearly identical results to much smaller coefficients. The Courant factor used is $\Delta x / c_{s} \Delta t = 0.5$, and the time step is set by the minimum required by any zone.  The hydro code has been verified via comparisons to the Sedov-Taylor and 1D shock tube problems, by verifying that pre-supernova stellar models with no explosions remain in hydrostatic equilibrium, and by comparing the solutions for de-pressurized models with analytic freefall solutions (Eq. \ref{eq:1}). 

The initial conditions are taken from a wide range of pre-supernova progenitor star models from \citet{woosleyetal2002}\footnote{http://homepages.spa.umn.edu/~alex/stellarevolution/data.shtml}. Three sets of models are considered: zero and solar metallicity progenitors with ZAMS masses of $11-40 M_\sun$, and $10^{-4} Z_\sun$ progenitors with ZAMS masses of $11-60 M_\sun$.  The lower mass solar metallicity progenitors retained their hydrogen envelopes and tended to be red supergiants ($R \sim 10^{14} \rm cm$), while those in the high mass range were bare helium or C/O stars ($R \sim 10^{11} \rm cm$). Low metallicity stars tended to be blue supergiants of smaller radii ($R \sim 10^{12-13} \rm cm$). Due to large uncertainties in prescriptions for semi-convection, convective overshoot, and mixing, we view the very low metallicity models as alternative outcomes for possible massive star progenitors rather than necessarily corresponding to extremely metal-poor environments.

Explosions are simulated using a moving inner boundary \citep[``piston,'' e.g.][]{woosleyweaver1995}. For the first $0.45$s the boundary moves inwards from the location where the specific entropy $s=4$ to $r = 5\times10^{7} \rm cm$, after which time it moves outwards at constant velocity. The inner boundary velocity is set to zero either after a specified amount of time or after the internal energy has changed by the desired amount. The resulting explosions are fairly insensitive to the piston velocity as long as it is large enough to deposit the desired amount of energy in a few seconds. We use the same number of radial zones for the hydrodynamics calculations as were used for the stellar evolution ($\simeq 500-1000$), but verify that doubling the number of zones and interpolating  using the nearest neighbor from the initial condition does not lead to significant changes in the evolution. 

An outflow boundary condition is employed by copying the acceleration from the outer zone to a single ghost zone. The inner boundary condition can be either a hard (reflective) piston or inflow, depending on the time.  Initially, the piston is used  to blow up the star, after which the inner boundary velocity is set to zero. To allow inflow, once the velocity of the inner zone drops below zero, it is copied to the inner boundary. This allows us  to use the inner boundary to blow up the star and to record the accretion rate once material begins to fall back. When the inner zone passes through the radius corresponding to the assumed outer disk edge, $r_{\rm min} = 10^8$ cm, its properties are saved and it is removed from the calculation. The outside of the accreted zone then becomes the inner boundary for the subsequent evolution.

We show  in Figure \ref{pvd} an example of the $1.2 \times 10^{51}$ erg explosion of a $25 M_\sun$, zero metallicity stellar progenitor. In this star, a significant density discontinuity at the helium/hydrogen interface ($r \simeq 2\times10^{10}$ cm) as well as the compact hydrogen envelope ($\rho \propto r^{-2.5}$) lead to a strong reverse shock forming at $t \simeq 20$ s. The two-shock structure can be seen clearly in the curves of $\rho (r)$ and $v (r)$. The filled circles show the results using a pure piston inner boundary condition, while the solid lines show the results for the ``fallback" boundary condition (i.e., piston switched to inflow after the shock was initiated).  The two are in excellent agreement in the portion of the star with $v > 0$, but differ slightly in the inner regions, as pressure support  slows the infall in the pure piston model. In the fallback calculation, the reverse shock turns around and leads to a jump in the accretion rate at $t \simeq 2000$ s. This calculation is very similar to that shown in Figure 1 of \citet{Zhang:2008p9515}, and the solutions are in qualitative agreement. The quantitative differences are likely due to a difference in progenitor models.

We follow the explosions until late time ($t = 10^8$ s), and calculate the accretion rate through the inner boundary from the properties of accreted zones. Sample accretion rate curves are shown in Figure \ref{mdot}.  In some cases, we find large accretion rates $(\ga 10^{-6}~M_\odot~{\rm s^{-1}})$ for a week or so after the explosion.  This late time accretion is due either to the fallback of stellar layers at large radii, or from the deceleration of inner layers by the reverse shock.  The energy associated with accretion at these rates is sufficient to power luminous supernova light curves.

\subsection{Semi-Analytic Treatment}

The general behavior of the fallback accretion rate can be easily understood  in the two limits where the material is either highly bound or mildly bound.  For the highly bound material (i.e., those layers where the velocity following the shock propagation is much less than the escape velocity) the accretion rate can be estimated from the free-fall time, 

\begin{equation}
t_{\rm ff} = \frac{\pi r^{3/2}}{\sqrt{8 G M(r)}},
\end{equation}

\noindent from each radial and mass coordinate in the progenitor star:

\begin{equation}\label{eq:1}
\dot{M} \equiv \frac{dM}{dr} \frac{dr}{dt} = 4\pi \rho(r) r^2 \frac{dr}{dt_{\rm ff}}.
\end{equation}

For an approximately power law density profile in a particular shell of a star, $\rho(r)=\rho_0 (r/r_0)^{3-\alpha}$, the enclosed mass is $M(r)=\rho_0 (r_0)\hspace{2pt} r_0^3\hspace{2pt} (r/r_0)^{\alpha}$ (for $0 < \alpha < 3$), and the fallback accretion rate is:

\begin{equation}
\dot{M}=\frac{8\pi\alpha}{3-\alpha}\frac{\rho_0 r_0^3}{t_0} \left(\frac{t}{t_0}\right)^{3(\alpha-1)/(3-\alpha)},
\end{equation}

\noindent where $t_0 \equiv (2G\rho_0)^{-1/2}$ (cf. Eq. 2 of QK12). For $\alpha < 0$, the enclosed mass is roughly constant, and the accretion rate is: 

\begin{equation}
\dot{M}=\frac{8\pi}{3} \frac{\rho_0 r_0^3}{t_0} \left(\frac{t}{t_0}\right)^{(2\alpha-3)/3},
\end{equation}

\noindent where now $t_0 \equiv \pi r_0^{3/2} / \sqrt{2GM(r_0)}$. In this way, the freefall accretion rate is set by the density profile of the progenitor star.

For the other limit of mildy bound material with $v_{\rm esc} \simeq v$, the maximum radius, $r_1 \equiv r_0 (1-v^2/v_{\rm esc}^2)^{-1}$, becomes much larger than the initial one, $r_0$. Then the asymptotic fallback rate, $\dot{M} \propto t^{-5/3}$, applies \citep{Michel:1988p19164,Chevalier:1989p9522}. This asymptotic scaling applies at the latest times in all three curves in Figure \ref{mdot}. 

Using the ballistics solution from \citet{Chevalier:1989p9522}, we can bridge these two asymptotic limits to analytically estimate the fallback accretion rate at all times for comparison with our numerical calculations.
For each mass shell, the downstream shock velocity is taken from the analytic formulae in \citet{Matzner:1999p20134}, which are  typically an excellent approximation to the numerical calculations. Then the total fallback time for each mass element can be calculated from Eq. 3.7 of \citet{Chevalier:1989p9522}, and its time derivative is an approximate accretion rate. This assumes that pressure effects are negligible, which is incorrect. However, the true acceleration measured from the numerical calculations described below turns out to be roughly constant at half of the gravitational acceleration.

This ballistic estimate reproduces the fallback accretion rate at all times in many progenitors. However, in some cases \citep[particularly blue supergiants such as SN1987A,][]{Chevalier:1989p9522} the reverse shock formed at the hydrogen-helium interface is strong enough to decelerate portions of the ejecta below the escape speed. This enhances the accretion rate at late times, and can significantly add to the remnant mass \citep{Zhang:2008p9515}. The reverse shock formation and evolution is analagous to that formed when the forward shock breaks out of the star and into the interstellar medium \citep[e.g.,][]{McKee:1974p19900,Chevalier:1982p17468}. As the simplest possible reverse shock prescription, we solve the strong shock jump conditions for the reverse shock velocity and the downstream velocity at the boundary of $100\%$ helium and hydrogen layers: $v_{RS} \simeq 0.6v_0$, where $v_0$ is the shock velocity. The reverse shock velocity evolves in time as the densities in both the expanding ejecta and unshocked hydrogen envelope change, and eventually it turns around. For simplicity, we ignore this and take $v_{RS}$ to be constant at its initial value. Then the location of intersection between ejecta and the reverse shock can be found, as well as the resulting ballistic $t(M)$ for material that is recaptured after passing through the reverse shock. The reverse shock prescription is important for the Z29 curve in Figure \ref{mdot}. This approximate semi-analytic description does a reasonable job reproducing the numerical calculations in all cases. The largest disagreement is in the reverse shock cases, where the semi-analytic accretion rate overestimates (underestimates) the numerical results at early (late) times. For the remainder of the paper, we use the results from the numerical fallback calculations. 

\begin{figure}
\epsscale{1.0}
\plotone{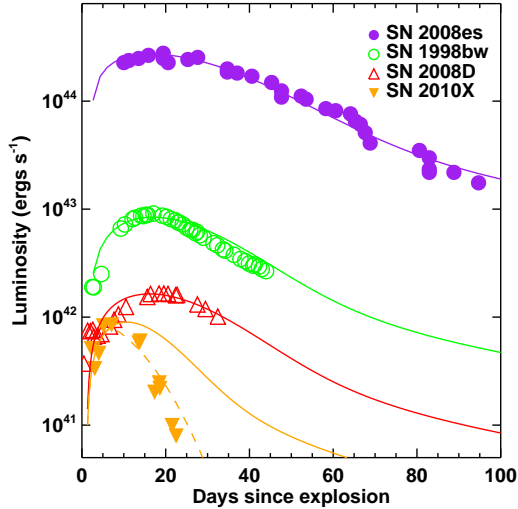}
\caption{\label{events}Comparison of fallback powered light curves (solid lines) from models U45 (purple), S33 (green), S39 (red), and S39O (orange) with some observed supernovae. The parameters for these events are given in Table \ref{tab:events}. The orange dashed curve assumes $t_{\rm off} = 7$ days. Data points are taken from \citet[][SN 2008es]{gezarietal2009}, \citet[][SN 1998bw and SN 2008D]{mazzalietal2008}, and \citet[][SN 2010X]{kasliwaletal2010}.}
\end{figure}

\begin{figure}
\epsscale{1.0}
\plotone{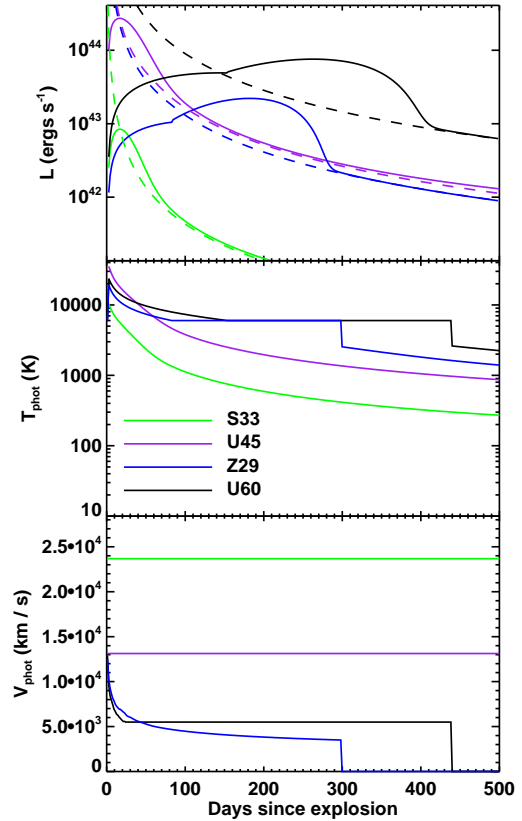}
\caption{\label{lcurves}Sample fallback powered light curves (top), and photospheric temperatures (middle) and velocities (bottom). The dashed curves in the top panel show the rate of energy injection from fallback accretion. The temperature is estimated from the one zone model, while the velocity is taken to be the maximum of $v_f$ and that calculated from the expanding ejecta. The temperature remains fixed at $T_I$ during the plateau phase for events where hydrogen is present.}
\end{figure}

\section{Possible Outcomes}
\label{outcomes}

We detail here the possible outcomes of supernova light curves powered by accretion energy. We first assume that a supernova  explodes via the traditional core collapse mechanism, whatever that may be.  For each progenitor, we ran explosions with a variety of energies, in the range $10^{48}$ to $10^{51}$~ergs,  in order to explore the full range of possible outcomes. Only explosions with positive total energy of non-accreted material at $t=10^8$ are considered, and the resulting remnant vs. initial mass distribution from these explosions is in excellent agreement with \citet{Zhang:2008p9515}.

The ejection of some stellar layers and the fallback of others is then calculated numerically as described in \S\ref{sec:numer-calc}, which determines the energy input rate from fallback. We then calculate approximate one zone light curves using the methods described in Appendix \ref{sec:light-curve-modeling}. For these calculations, we need the effective diffusion time through homologously expanding ejecta \cite{Arnett_1979},

\begin{equation}
\label{diffusion}
t_d = \sqrt{\frac{3}{4\pi}\frac{M \kappa}{v c}} = \sqrt{\frac{3}{4\pi} \frac{(M_{\rm ej}+M_{\rm fb})\kappa}{v_f c}},
\end{equation}

\noindent where $M_{\rm fb} = \xi \int \dot{M}_{\rm fb} dt$ is the total outflow mass, $E_{\rm fb} = \epsilon \dot{M}_{\rm fb} c^2$ is the injected accretion energy, and $v_f = \sqrt{(E_{\rm sn}+E_{\rm fb})/(M_{\rm ej}+M_{\rm fb})}$ is the final ejecta velocity. Note that there is an ambiguity in determining $M_{\rm fb}$, depending on the interpretation of the fudge factor $\xi$. If $\xi$ indicates the fraction of outflow mass that interacts with the supernova ejecta, then the above expression for $M_{\rm fb}$ applies. If on the other hand, the mass transfer to the ejecta is more efficient while the specific energy of the outflow is lower, $M_{\rm fb}$ could be significantly larger.

We assume a constant opacity $\kappa=0.2 \hspace{5pt} \rm g \hspace{5pt} \rm cm^{-1}$, appropriate for electron scattering for fully ionized elements heavier than hydrogen. This is clearly a coarse approximation, as the actual opacity will depend on the composition and the presence of Doppler broadened lines.  The effects of recombination on the opacity are, however, included in an approximate way (Appendix \ref{sec:light-curve-modeling}).  

While our one zone light curve models account for the acceleration of the ejecta due to the input accretion energy, they lack any information on the radial structure of the ejecta.  The radiation hydrodynamical calculations of KB10 show that energy deposition at the base of the ejecta (in that case from a magnetar) blows a bubble in the inner regions, piling up material into a dense shell.  We expect a similar effect in fallback powered SNe, which will likely also induce an asymmetry if the energy deposition is anisotropic.

For each of the light curves, we measure the time to peak, $t_p$, and the peak luminosity, $L_p$. The results are shown in Figure  \ref{lptpl} for $\epsilon=10^{-3}$. Each point represents a single explosion energy and progenitor model, color-coded by the radius of the pre-supernova star: red for $R > 10^{13} \rm cm$ (RSGs), purple for $10^{12} \rm cm < R < 10^{13} \rm cm$, blue for $10^{11} \rm cm < R < 10^{12} \rm cm$ (BSGs), and green for $R < 10^{11} \rm cm$ (He or C/O stars). This radius also corresponds to the zero age main sequence metallicity: solar without significant mass loss for RSGs, zero for BSGs, $10^{-4} Z_\sun$ for stars in between, and solar with large amounts of mass loss for compact He and C/O stars. Events are only plotted if $L_p$ is larger than the thermal supernova luminosity, 

\begin{equation}
\label{eq:3}
L_{\rm sn} \sim \frac{E_0}{t_d} \frac{R}{v t_d}.
\end{equation} 

\noindent The number of points is then set by the number of explosion energies and progenitor models, as well as the fraction of cases where that condition is met. The number of points does not represent an expected rate, since both the choices of explosion energies and progenitor models are arbitrary.

Figure  \ref{lptpl} illustrates the wide range of light curves that may result 
when fallback power is included.
Many of the successful explosions with  energies $\sim 10^{51}~ \rm ergs$ lead to events with $t_p \simeq 50-200$ days, $L_p \sim 10^{41-44} \es$. The long durations are similar to those of Type~II plateau SNe, and a result of the large ejecta masses and correspondingly long diffusion times.  The final velocities of these events are also fairly typical of core-collapse supernova explosions ($\sim 3000~ \rm km \rm s^{-1}$). This is because the amount of fallback is much less than the ejecta mass, so that fallback energy does not appreciably change the total kinetic energy of the explosion. For smaller ejecta masses, the fallback energy can dominate the total explosion energy, significantly increasing the final velocity. The diffusion timescale therefore decreases with decreasing ejecta mass both from the smaller total mass and because of the increasing final ejecta velocity. 

These effects lead to a strong scaling of $L_p$ with $v_f$, shown in Figure \ref{tpvf}. The roughly $L_p \propto v_f^2$ dependence can be recovered by assuming the fallback energy always dominates the supernova energy ($E_{\rm fb} \sim v_{\rm f}^2$), while the fallback mass contributes negligibly to the ejecta mass. Furthermore, the scaling assumes that the total fallback energy scales with peak luminosity ($L_p \sim E_{\rm fb} \sim v_f^2$), which is true if the accretion rate at late times scales with its integral over all times. The apparent maximum in $v_f \lesssim 3 \times 10^{9} \rm cm \rm s^{-1}$ is from the case where the fallback mass and energy dominate that of the supernova explosion: $v_f \simeq \sqrt{\epsilon / \xi} c = 0.01 c$ for our standard parameters. In the context of the simple outflow models described in Section \ref{accenergy}, this maximum velocity scales as $v_{f,\rm max} \propto r_{\rm out}^{-1/2}$. The considerable scatter in Figure \ref{tpvf} is from the breakdown of the above assumptions. 

\subsection{Candidate Events}
\label{sec:candidate-events}

Different classes of progenitor stars lead to different outcomes in Figure \ref{lptpl}. First, solar metallicity RSG progenitors for the most part lead to relatively low luminosity events ($L_p \lesssim 10^{43} \es$). At high ZAMS masses, these stars undergo substantial mass loss and become stripped He or C/O stars. These progenitors can lead to events with $t_p \simeq 20 \rm days$, $L_p \sim 10^{42-43} \es$. These could potentially explain broad line Type Ibc GRB SNe: high velocities are a natural outcome of the injection of large amounts of fallback energy. Example fits are shown in Figure \ref{events} for SN 1998bw \citep{galamaetal1998} and SN 2008D \citep{soderbergetal2008}. In the context of the collapsar model, this suggests that the central engine could be responsible for all of the observed properties: early time accretion leading to black hole formation, the GRB, and the initial supernova explosion; and late time accretion powering the resulting light curve and the large expansion velocities. 

BSG progenitors lead to two classes of outcomes depending on the explosion energy. At low explosion energies, they can produce luminosities $L_p \sim 10^{43-45} \es$ and peak times of $t_p \sim 2-40$ days. The short durations are from the very small ejecta masses, $M_{\rm ej} \sim 10^{-3}-10^{0} M_\sun$, with nearly all of the star falling back. For $\xi = 0.1$ used here, the wind mass is comparable to the ejecta mass, and the injected fallback energy is much larger than the initial explosion energy. This leads to large final velocities and short diffusion times. Events with $t_p \sim 25$ days can have light curve shapes very similar to observed luminous Type II-L events. An example fit to the superluminous Type II-L SN 2008es \citep{gezarietal2009} is shown in Figure \ref{events}.

At high explosion energies, BSG progenitors lead to a range of long duration events with $t_p \simeq 100-300$ days, $L_p \sim 10^{42-44} \es$. The most luminous cases are either from very massive stars ($\gtrsim 40 M_\sun$) at low metallicity or from zero metallicity stars with strong reverse shocks. In both cases, the ejecta masses are $\simeq 10-40 M_\sun$ with low expansion velocities, $v_f \simeq 2000-6000 \rm km \rm s^{-1}$.

Subluminous Type I and II events are possible on a variety of timescales. As an example, Figure \ref{events} shows a comparison of a Type I explosion with the transient 2010X \citep{kasliwaletal2010}. The steep decay in this case requires that the accretion turn off about 7 days after explosion (see \S\ref{interactejecta}). 

Approximate light curves from examples of each of these type of events are shown in Figure \ref{lcurves} along with photospheric temperatures and velocities.  The model parameters are listed in Table \ref{tab:events}. The photospheric temperature is taken from the one zone light curve calculations (see Appendix \ref{sec:light-curve-modeling}). The photospheric velocity is taken to be the maximum of $v_f$ and the photospheric velocity in the expanding ejecta in the absence of injected accretion energy. For light curves with recombination, the photospheric properties are meaningless after the plateau phase, since then formally the ejecta are completely optically thin. For relatively short events, the expansion velocities are high ($\gtrsim 10^9 \rm cm \rm s^{-1}$, and the fallback energy sets the velocity since the ejecta mass is small ($\simeq 1-2 M_\sun$). Much slower velocities occur in the longer duration events with large ejecta masses ($\simeq 10-40 M_\sun$). The photospheric temperatures are very high at peak in II-L type events ($\simeq 20000 K$).

When recombination isn't important, the light curves are in excellent agreement with the semi-analytic formula in Eq. (\ref{powerlawlc}) for a power-law injection of energy with $n=5/3$. This is because in nearly all cases the late time accretion rate falls as $\dot{M} \propto t^{-5/3}$, while any energy injected on timescales $\lesssim 1$ day is lost to adiabatic expansion, so that its time-dependence does not influence the light curve.

\section{Caveats}

Gravitational energy liberated through fallback accretion at late times can power unusual supernova light curves (Figures \ref{events} and \ref{lcurves}). 
The calculations in this paper have made many simplifying assumptions; we discuss here some of the uncertainties.  We have treated the explosion of stars with crude 1D hydrodynamic calculations using a piston. This method has frequently been used to simulate core collapse supernova explosions and the resulting fallback \citep[e.g.,][]{woosleyweaver1995,MacFadyen:2001p9343,Zhang:2008p9515}, and the uncertainties in the numerically calculated fallback accretion rates are probably less than those in the outflow physics and/or parameters (e.g., $\epsilon$). The light curve calculations further assume simple one zone prescriptions for the bolometric luminosity and photospheric temperature. More sophisticated techniques would be required for spectral calculations.

The density structures of the pre-supernova models, which directly impact the fallback rate, depend sensitively on uncertain prescriptions for convection (semi-convection and overshoot) and compositional mixing in stellar evolution calculations \citep[e.g.,][]{woosleyetal2002}. Probably a bigger issue is that the calculations here are based on a limited set of stellar progenitors, and ignore the effects of rotation and binarity, which may be very common in massive stars \citep[e.g.,][]{deminketal2012}. There may be additional variety in the range of possible fallback powered transients from stellar progenitors not considered here. 

Further, we have assumed that the stellar material that falls back after the explosion has sufficient angular momentum to form a disk, and that this disk can efficiently drive a massive wind and/or ultrarelativistic jet. These are both important open questions. The angular momentum distribution and surface rotation rates of massive stars remain highly uncertain \citep{Woosley:2011p19913}. Although previous studies have found prominent polar outflows from geometrically thick black hole accretion flows \citep{Stone:1999p19169,igumenshchevabramowicz2000}, more recent calculations have found large-scale circulations to be more common than unbound massive winds \citep{mckinneyetal2012,narayanetal2012}. If so, ultrarelativistic jets may be a more natural explanation for injecting energy into the ejecta. 

Finally, we have assumed that this outflow can thermalize in the outgoing supernova ejecta without expelling infalling material and halting accretion. We outline the requirements below to satisfy these assumptions, and estimate in a few sample cases the required rotation rates for disk formation.

\begin{figure}
\includegraphics[scale=0.9]{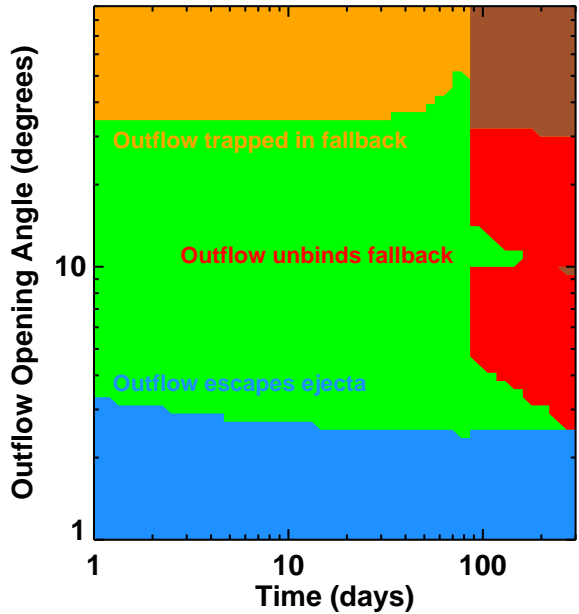}
\caption{\label{tu45}Regions of $\theta_j$ vs. $t$ parameter space for model Z29 where i) the outflow cannot escape the accreting material before depositing most of its energy (Eq. \ref{tfb}), ii) the outflow escapes the outgoing supernova ejecta before losing most of its energy (Eq. \ref{tej}), and iii) the energy deposited in the accreting material by the outflow exceeds its binding energy (Eq. \ref{edep}). The remaining parameter space is where an outflow could plausibly power a supernova light curve without shutting off continuing accretion. For $\theta_j < 10^\circ$, the outflow is arbitrarily changed from a wind ($v_j = 0.1 c$) to an ultrarelativistic jet ($v_j \simeq c$). Constraints i) and ii) fix the range of allowed $\theta_j$ for any disk formation time, $t_{\rm on}$, while constraint iii) sets the time at which fallback accretion will stop ($t_{\rm off}$).}
\end{figure}

\begin{figure}
\epsscale{1.0}
\plotone{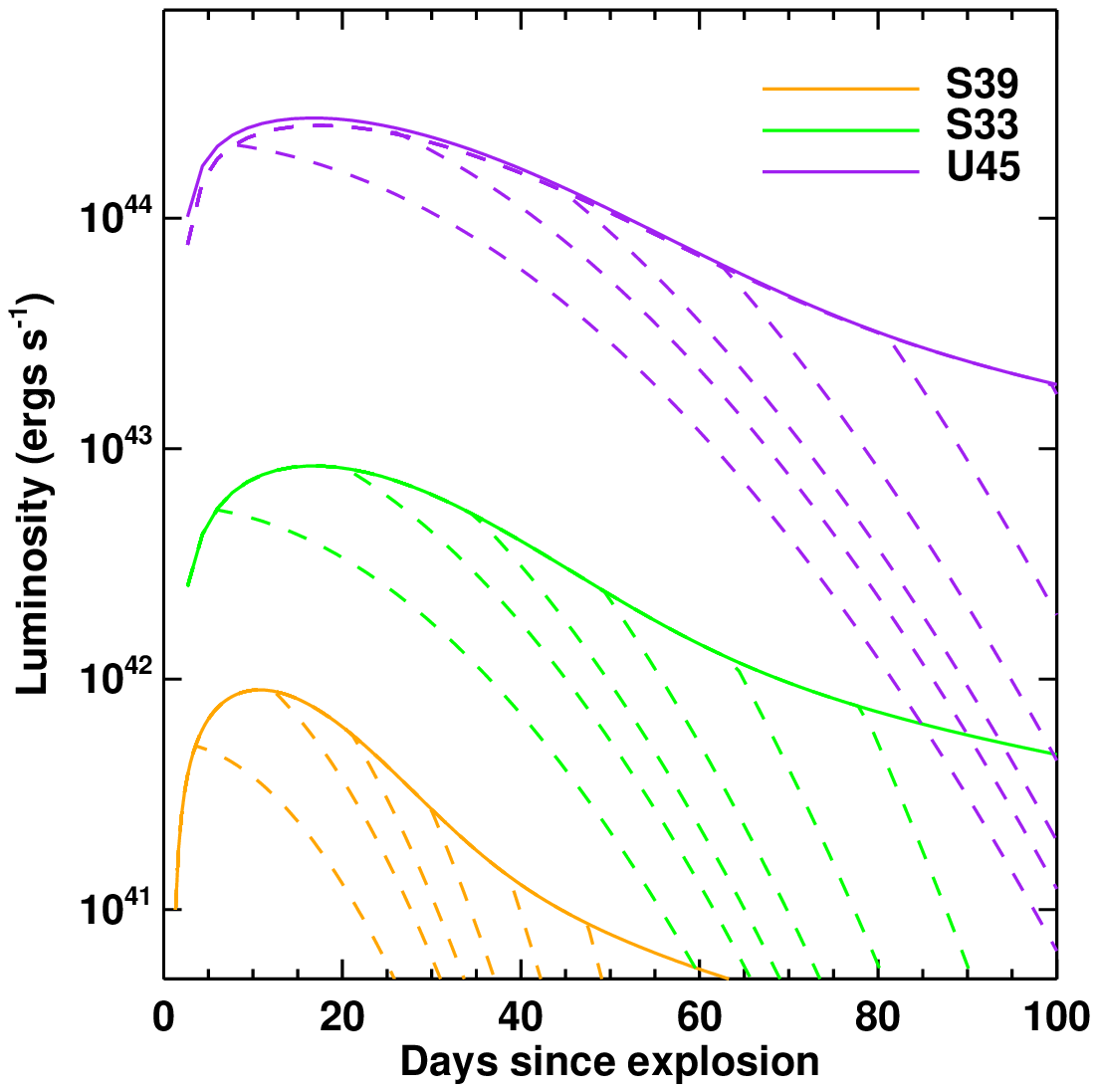}\\
\plotone{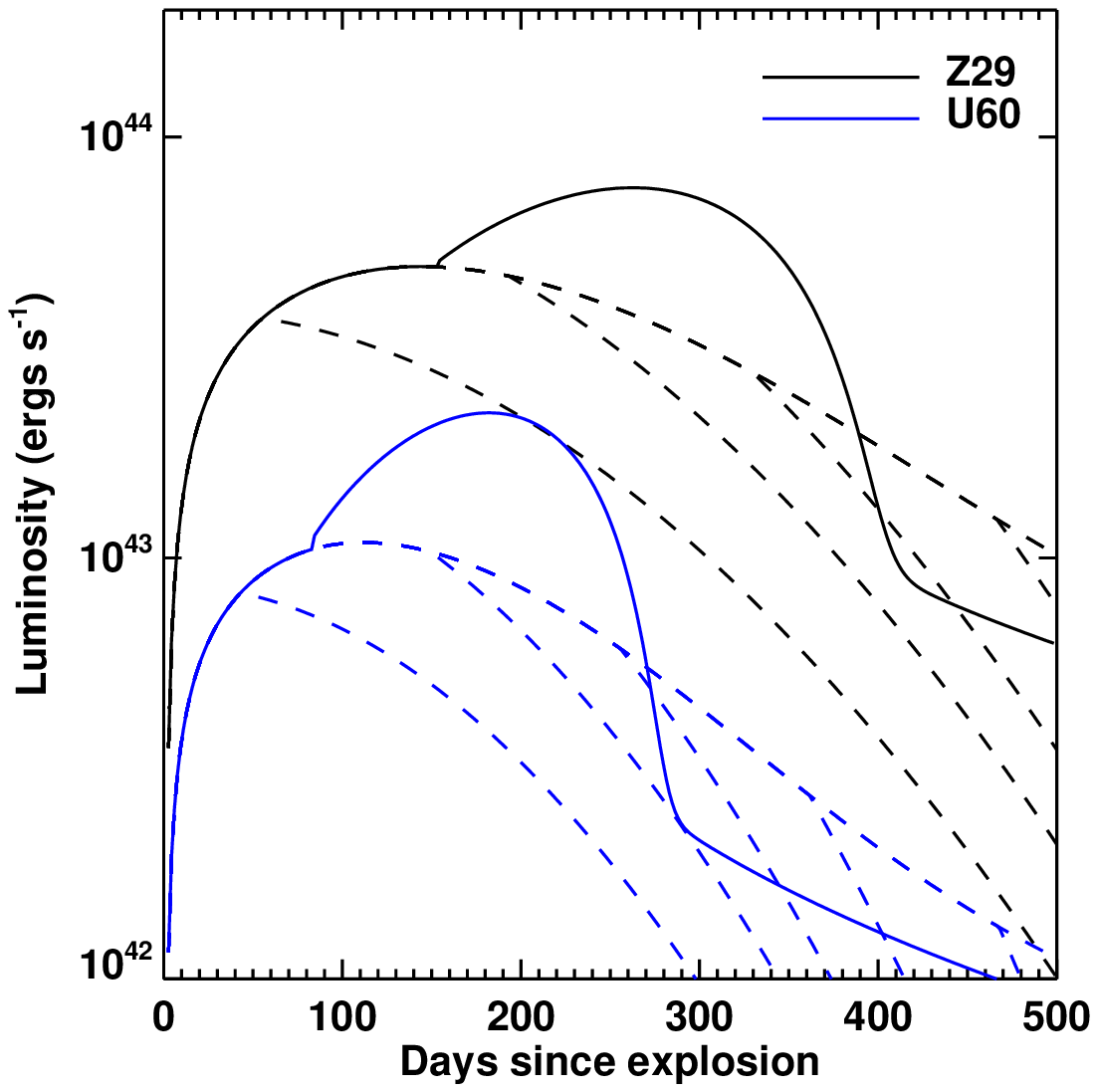}
\caption{\label{lcurveonoff}Sample fallback powered light curves (lines) for models (top panel) U45, S33, S39O; and (bottom panel) U60 and Z29. The solid curves assume continued energy injection, while the dashed curves turn off at a range of times, $t_{\rm off} = 0.3-4.0 t_d$. The dashed curves all assume constant opacity, and the strong effects of recombination can be seen on the light curves in the bottom panel.}
\end{figure}

\begin{figure}
\epsscale{1.0}
\plotone{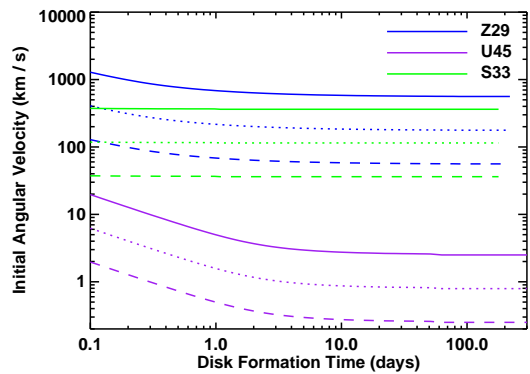}
\caption{\label{vstar}Angular velocities required to form disks at radii at $r=10^8 \rm cm$ (solid), $10^7 \rm cm$ (dotted), and $10^6 \rm cm$ (dashed) in a few models from material falling back at a range of turn on times.}
\end{figure}

\subsection{Outflow Collimation and Interaction with Ejecta}
\label{interactejecta}

In order for a fallback accretion powered outflow (either a ultrarelativistic jet with speed $v_j \simeq c$ or massive wind with $v_j \sim 0.1 c$) to power a supernova light curve, it must be able to: i) escape the remaining infalling material ii) without unbinding it and iii) thermalize in the outgoing supernova ejecta. To order of magnitude, we assess the plausibility for this scenario as follows.

Following QK12, we assume a magnetically-dominated outflow, collimated with an opening angle $\theta_j$. The propagation of the outflow through the remaining bound fallback is similar to the propagation of a jet through a host star during a long GRB \citep[e.g.,][]{begelmancioffi1989,matzner2003}. The speed of the head of the collimated outflow, $v_h$, is determined by pressure balance between the outflow and the host star (Eq. 4 of QK12):

\begin{equation}
  v_h \simeq \left(\frac{4 L_j R}{3 M v_j \theta_j^2}\right)^{1/2},
\end{equation}

\noindent where $R$ is the maximum radius of bound material and $M$ its total mass. The outflow escape timescale is then $t_{\rm esc} \sim R / v_h$. The outflow also drives a lateral shock into the surrounding material, whose speed is approximately (QK12),

\begin{equation}
  v_l \simeq f^{1/4} \theta_j^{1/2} v_j^{1/4} v_h^{3/4},
\end{equation}

\noindent where $f$ is the efficiency of depositing outflow energy in the surrounding material. If the outflow is dominated by toroidal magnetic field (e.g., in a helical jet or outflow from a rotating disk), a typical value from numerical calculations is $f \simeq 0.03$ \citep{Bucciantini:2007p3784}. For the outflow to escape, $t_{\rm esc}$ should be shorter than the time for the lateral shock to envelope the star, $t_{l} \sim R / v_l$. For interactions with the remaining bound fallback material, this ratio is:

\begin{equation}\label{tfb}
\frac{t_{l}}{t_{\rm esc}} \simeq 0.6 \left(\frac{M_{\rm fb}  v_j^3 \theta_j^6}{L_j R_{\rm fb}}\right)^{1/8}.
\end{equation}

In addition to requiring $t_l > t_{\rm esc}$, continued accretion requires that the energy deposited, 

\begin{equation}\label{edep}
E_{\rm dep} = f \int_{t_{\rm on}}^{t_{\rm on} + t_{\rm esc}} L_j (t) dt,
\end{equation}

\noindent should be less than the binding energy of the remaining fallback material, where $t_{\rm on}$ ($t_{\rm off}$) is the time after explosion at which the outflow turns on (off). 

Conversely, for the outflow energy to be deposited efficiently in the ejecta the outflow escape time should be shorter than the energy deposition time. Since the mechanism for thermalizing the outflow energy and its associated timescale are unknown, we can instead use the same comparison of $t_l$ and $t_{\rm esc}$ as above. 

In this case, $v_{\rm ej}$ is larger than $v_h$, and the escape time can be estimated from setting $R = v_{\rm ej} t$ and finding when $v_h = v_{\rm ej}$ (QK12):

\begin{equation}
t_{\rm esc} = \frac{3 M_{\rm ej}  v_{\rm ej} v_{\rm j} \theta_j^2}{4 L_j}.
\end{equation}

Similarly, we can find the ratio $t_l / t_{\rm esc}$ under the same assumptions. The result is:

\begin{equation}\label{tej}
\frac{t_l}{t_{\rm esc}} \simeq 20 \left(\frac{f \theta_j^2 v_j}{v_{\rm ej}}\right)^{2/3}.
\end{equation}

The requirements i) $t_{\rm esc, FB} / t_{l, \rm FB} < 1$, ii) $E_{\rm dep}/E_{\rm grav} < 1$, and iii) $t_{l, \rm ej} / t_{\rm esc, ej} < 1$ amount to constraints on the outflow opening angle, $\theta_j$, and the time over which fallback accretion can continue. Excluded regions of $\theta_j$ vs. $t$ parameter space from enforcing these constraints for the model Z29 are shown in Figure \ref{tu45}. 

For the interaction of accretion energy with remaining fallback material, we find the maximum radius reached at time $t$ by material that will ultimately accrete ($R_{\rm fb}$), and its remaining total mass ($M_{\rm fb}$) and binding energy ($E_{\rm grav}$). For the interaction of the accretion energy with the ejecta, we use $M_{\rm ej}$ and $v_{\rm f}$ estimated at time $t$. 

The timescale constraints essentially place limits on $\theta_j$ for each type of outflow for all disk formation times, $t_{\rm on}$: at small (large) opening angles, the outflow escapes (is captured). These ratios also depend on the other quantities, leading to differences between various models. Generally, smaller $v_f (t)$ leads to higher ejecta densities and help to trap the outflow. Outflows escaping the ejecta before thermalizing could appear as long duration, high energy transients \citep[QK12,][]{Woosley:2011p19913}. Outflows trapped in the material still falling back would likely deposit energy there more effectively, either unbinding the material or prolonging its accretion to later times.

At late times, the outflow will unbind any remaining material, shutting off further accretion. This is because the energy deposition into the accreting material at late times scales as $L t_{\rm esc} \sim t^{-5/6}$, while its binding energy scales as $M_{\rm fb} / r_{\rm fb} \sim t^{-4/3}$. Equating these gives the turn off time ($t_{\rm off}$) for each event. This turn off time tends to be shorter in higher energy explosions, since the bulk of the late time accretion comes from loosely bound material. Sample light curves for events where accretion shuts off are shown in Figure \ref{lcurveonoff} for a range of $t_{\rm off}$, assuming a constant opacity. Once the injected energy runs out, the light curve decays according to Eq. (\ref{freelc}), but with an initial luminosity $L( t_{\rm off})$. This may be particularly relevant for long duration transients in models like U60 and Z29, where the turn off time ($\simeq 80$ days for Z29) is likely to be comparable to the time to peak. 

Although these estimates demonstrate the plausibility of accretion-driven outflows powering supernova light curves, detailed physical calculations will be required to assess this scenario accurately. Further, the statement that the outflow cannot escape the ejecta does not provide an efficient means of thermalization, since we have assumed $f = 0.03$. Using a larger value of $f \approx 1$ would shift the range of allowed opening angles to favor relativistic jets, and lead to the outflow unbinding the accreting the material time at proportionally earlier times. 

The efficiency of thermalization depends on how exactly energy is transported from the accretion disk to the supernova ejecta. We have assumed that this mechanism is a highly magnetized disk wind or an ultra-relativistic jet. If instead the wind is not highly magnetized, a double (forward/reverse) shock structure will form when it catches up with the slowly moving inner layers of the supernova ejecta (KB10). The situation is analogous to the commonly case of supernovae interacting with circumstellar material,  only here the interaction happens inside, rather than outside the remnant. In either case, shocks should be efficient in thermalizing the kinetic energy of the wind. Some recent semi-analytic \citep{begelman2012} and numerical \citep{mckinneyetal2012,narayanetal2012} calculations of non-radiative accretion flows have found large-scale circulations or convective motions \citep{Lindner:2011p20292} as well as or instead of outflows. This may also be a relevant mechanism for transporting accretion energy to large radius.

If the accretion energy cannot efficiently thermalize, it will likely still lead to high ejecta velocities $\simeq v_f$. In the case of an event like S33, this could still explain broad line Type Ibc supernovae: radioactivity would power the light curve and accretion energy would lead to the high observed photospheric velocities. This is also a possible outcome of early time accretion onto a magnetar \citep{Piro:2011p17810}.

\subsection{Angular Momentum and Disk Formation}
\label{angmom}

Given the viable range of disk formation times for fallback accretion powered supernovae \ref{interactejecta} and the initial stellar radii accreting at those times, we can calculate the required angular velocity. Curves for models Z29, U45, and S33 are shown in Figure \ref{vstar} for forming disks at radii from $10^{6-8} \rm cm$, or $r \sim 1-100$ for a $10 M_\sun$ black hole. Naively assuming rigid rotation, in all cases disks can form at the required times without exceeding breakup at the outer edge of the star. The required rotation rates essentially scale with explosion energy: for large explosion energies, the envelope is expelled, and larger rotation rates are required for the disk to form from material that was originally at smaller radius. 

Under this assumption, we can also calculate the maximum disk size from fallback accretion, and the corresponding viscous time, $t_{\rm visc} \sim (R/H)^2 \alpha^{-1} t_{\rm dyn}$, where $H/R$ is the accretion flow scale height and $\alpha$ is the standard dimensionless viscosity parameter in accretion theory \citep{shaksun1973}. Even with conservative assumptions ($H/R = 0.1$, $\alpha = 0.01$), this timescale only becomes larger than the fallback timescale for the highest rotation rates and early disk formation times $\lesssim 10^4$s. This is because assuming rigid rotation, the total disk size never greatly exceeds its formation radius.

Stellar cores spin up as they contract, and depending on the efficiency of angular momentum transfer from magnetic torques \citep{spruit2002} can transfer much of the core angular momentum to the outer layers \citep[e.g.,][]{hegeretal2005}. If this mass is retained, as in our models from low metallicity progenitors (e.g., Z29 and U45), it will likely form a disk upon fallback for modest ZAMS rotation rates. If instead this mass is lost (e.g., S33), insufficient angular momentum may remain to form a disk. If the red supergiant is in a binary system, tidal interactions may be an efficient means to spin up the star sufficiently to cause disk formation even if the envelope is lost during subsequent evolution \citep{Woosley:2011p19913}. The latter scenario may be fairly common, given the frequency of massive stars in binaries \citep{deminketal2012}. These scenarios should be considered in more detail in future work.

\section{Conclusions}
The accretion power released when material falls back onto a compact remnant at late times could power unusual supernova light curves. We have explored the consequences for a variety of progenitors and explosion energies, using numerical calculations of the fallback accretion rate and order of magnitude estimates of the resulting energy injection. While most of the fallback typically occurs at early times, it may be significant at late times in very massive stars, for low explosion energies, or when a strong reverse shock forms at the hydrogen/helium boundary. We have demonstrated that it is plausible that, under certain circumstances, the energy available from accretion could power an outflow which then thermalizes in the supernova ejecta.  

The events we have described are different and more diverse than what have previously been studied as ``fallback supernovae''.  \citet{fryer2009a}, for example, considered the case of massive star collapse in which most of the material fell into the central black hole and only a fraction was ejected.  Because they also assumed that the surrounding medium was very dense and extended (due to mass loss prior to explosion) the supernova shock wave did not breakout of the circumstellar gas until late times.  The result was a dim, shock-powered transient lasting from weeks to months.  \citet{moriya2012}  similarly considered the case in which most of the star fell back and only a very small amount ($\sim 0.1 M_\odot$) was ejected.  By assuming that this ejecta was enriched with $^{56}$Ni,  they found a brief and sub-luminous radioactively powered transient similar to SN~2005E. Both of these  previous scenarios neglected the possible input of accretion energy from fallback (i.e., they assumed $\epsilon=0$).  As we have shown,  accretion may re-energize the ejecta at late times and hence power much brighter emission.

The power from fallback accretion may be relevant for explaining recently discovered classes of peculiar supernovae.  These may include the Type~IIL supernovae that are extremely luminous and of relatively short duration \citep[e.g.,][]{gezarietal2009,miller2009} as well as those that are moderately bright and of very long duration  \citep[e.g.,][]{miller2010,rest2011,Chatz2011}.  Several of the observed Type~II events, however, also show narrow hydrogen emission features in their spectra, indicating that interaction with a dense circumstellar medium is occurring and may be responsible for the luminosity. 

Many of the predicted Type II events with $\sim 10^{51}$ erg explosion energies have very long times to peak ($100-200$ days). Accretion energy is likely to unbind the remaining infalling material on a comparable timescale, turning off the power source for the light curve (\S \ref{interactejecta}). We therefore predict that these events could be seen as very bright Type II supernovae that disappear suddenly from view. In general, late time turn off would be an observational signature of fallback accretion powered supernovae.

Fallback accretion could also power very bright Type~I events.  The models considered in this paper reached peak luminosities of $\sim 10^{43}~{\rm ergs~s^{-1}}$, similar to the broad-lined SNe~Ic like SN~1998bw.  If the accretion efficiency is assumed to be higher than our fiducial case, it is possible for some events to reach luminosities $\ga 10^{44}~{\rm ergs~s^{-1}}$, in which case fallback could power the super-luminous hydrogen poor events such as SN~2005ap \citep{quimbyetal2011}. 

Another effect that may produce super-luminous events like SN~2005ap involves mass loss.  Some events considered here (e.g., Z29) are brightened considerably by enhanced accretion from material decelerated by a reverse shock forming at the H/He interface. A similar outcome could occur in both Type I/II events where the progenitor has experienced considerable mass loss shortly before explosion. In this case, the reverse shock would be formed when the outgoing shock wave reaches the interface between the progenitor star and the massive wind or ejected shell.  The subsequent inward propagation of the reverse shock could lead to order of magnitude increases in the fallback accretion rate at later times. Interaction of supernova ejecta with circumstellar shells at radii $\sim 10^{15}$~cm is commonly considered to explain superluminous supernovae via thermalization of the kinetic energy \citep[e.g.,][]{galyam2012}.  Surprisingly, interaction  with circumstellar material at much smaller radii ($\sim 10^{11}-10^{12}$~cm) may also lead to an super-luminous event, but by very different means -- by enhancing fallback accretion and feeding the central compact object at late times.

Also of potential relevance to the fallback scenario are dimmer supernovae that decline very rapidly after peak \citep[e.g., SN~2002bj and SN~2010X,][] {poz2010,kasliwaletal2010}.  The short duration of these events makes it difficult to explain them as radioactively powered transients.  In the case of SN~2002bj at least, the mass of $^{56}$Ni inferred from the light curve peak exceeds the total ejecta mass inferred from the light curve duration (diffusion time), seemingly ruling out a radioactively powered events.  In the fallback scenario, short lived transients are possible, especially if the energy injection from accretion cuts off the fallback abruptly (Figure~\ref{lcurveonoff}).

The true range of possible light curves powered by fallback accretion is likely much larger than shown in Figure \ref{lptpl}. Those events are limited in both the variety of progenitor models and by our neglect of the fallback accretion physics. The latter could conceivably lead to variations in $\epsilon$ in either direction: smaller disks and/or more efficient thermalization of the outflow could lead to higher peak luminosities $\gtrsim 10^{45} \es$. Conversely, lower efficiencies could help explain a wider variety of sub-luminous supernovae \citep[e.g., SN 2008ha,][]{foleyetal2009}.

Further modeling is needed to identify the observational signatures of fallback accretion powered supernovae, and determine how we might distinguish these events from other means of generating unusual light curves. Possible signatures include a tail with $L \propto t^{-5/3}$ at late times compared to the diffusion time (but before the ejecta become completely optically thin), somewhat different from those from radioactivity or magnetar spindown. More promisingly, at late times ($\sim 100$ days) it seems likely that the accretion energy will unbind the infalling material (\S~\ref{interactejecta}), shutting off accretion and leading to a sudden decrease in luminosity. If instead accretion continues for decades after the explosion, the black hole could emerge as an observable X-ray source \citep{balbergetal2000}, as has been suggested for SN 1979C \citep{patnaudeetal2011}.

The conditions for fallback to influence the supernova light curve are apparently quite special, as the scenario requires sufficient angular momentum to form a disk and an evolution that permits fallback to persist long enough to drive energetic outflows at late times.    Such a confluence of factors may be rare in the Universe.  On the other hand, observational surveys show that the rate of peculiar SNe -- in particular the rate of very luminous ones -- is only a small fraction that of standard  core collapse events. It is possible that fallback power plays a role in some of these spectacular events.

\label{summary}

\begin{acknowledgements}

We thank A. Heger for making a large number of pre-supernova stellar models publicly available. JD thanks L. Bildsten, B. Metzger, C. Ott, T. Piro, E. Quataert, E. Ramirez-Ruiz, and S. Woosley for stimulating discussions related to this work. This work is supported  by the Director, Office of Energy Research, Office of High Energy and Nuclear Physics, Divisions of Nuclear Physics, of the U.S. Department of Energy under Contract No. DE-AC02-05CH11231, and by a Department of Energy Office of Nuclear Physics Early Career Award.

\end{acknowledgements}

\bibliographystyle{apj}

\appendix
\section{Light Curve Modeling}

\label{sec:light-curve-modeling}

KB10 described a one zone diffusion estimate for bolometric supernova light curves  powered by an injection of energy with arbitrary time-dependence, $H (t)$.   The argument follows along the lines of \citet{Arnett_1979, Arnett_1982}. As the ejecta expand, energy is lost both due to escaping radiation ($L$) and adiabatic losses from expansion:

\begin{equation}
\label{eq:2}
\frac{\partial (E_{\rm int} t)}{\partial t} = -p \frac{\partial V}{\partial t} + H - L
\end{equation}

\noindent Assuming optical depth $\tau \gg 1$, the diffusion equation gives an approximate relationship between $E_{\rm int}$ and $L$:

\begin{equation}
 F=\frac{L}{4\pi R^2}\approx\frac{c}{3 \kappa \rho} \frac{E_{\rm int} / V}{R}. 
\end{equation}

\noindent With the definition of the diffusion time (Eq. \ref{diffusion}) and assuming $V \propto t^3$, Eq. (\ref{eq:2}) can be re-written as: 

\begin{equation}
\left(\frac{d}{dt}+\frac{t}{t_d^2}\right) L (t) = \frac{t}{t_d^2} H (t).
\end{equation}

\noindent The general solution for $L (0) = 0$ is: 

\begin{equation}
\label{lcurve}
L (t) = \frac{e^{-t^2/2 t_d^2}}{t_d^2} \int \hspace{2pt} dt\hspace{2pt} t\hspace{2pt} e^{t^2/2t_d^2}\hspace{2pt} H (t).
\end{equation}

\noindent For an initial shock energy $H = E_0 \delta(t-t_0)$, this gives:

\begin{equation}\label{freelc}
L (t) = \frac{E_0}{t_d}\frac{t_0}{t_d} e^{-(t^2-t_0^2)/2 t_d^2},
\end{equation}

\noindent in agreement with Eq. (\ref{eq:3}) with $t_0 = R / v_{\rm sh}$ modulo the exponential factor of order unity at peak. 

For accretion powered light curves, a power law form, $H = L_0 (t / t_0)^{-n}$ with $n=5/3$ provides an excellent approximate description for the numerical light curves integrated with Eq. (\ref{lcurve}). The semi-analytic solution for $t > t_0$ is:

\begin{equation}
\label{powerlawlc}
L (t) = L_0 \left(\frac{t_0}{t_d}\right)^n e^{-t^2/2 t_d^2}  \left(-\frac{1}{2}\right)^{n/2} \left[\gamma\left(1-\frac{n}{2}, -\frac{t_0^2}{2 t_d^2}\right)-\gamma\left(1-\frac{n}{2},-\frac{t^2}{2 t_d^2}\right)\right],
\end{equation}

\noindent where $\gamma(s,x)$ is the lower incomplete Gamma function. The incomplete Gamma function is complex for negative arguments. Since the observed light curve and the integral in Eq. (\ref{lcurve}) are real, the imaginary part in Eq. (\ref{powerlawlc}) vanishes. In the special case of constant energy injection ($n=0$), the solution is (cf. Eq. 13 of KB10):

\begin{equation}
L (t) = L_0 \left[1-e^{-(t^2-t_0^2)/2t_d^2}\right],
\end{equation}

\noindent for $t \le t_{\rm off}$, where $H=0$ for $t > t_{\rm off}$.

This light curve estimate assumes a constant opacity. A different limit occurs when the outer portion of the ejecta drops below the ionization temperature and recombines. The opacity drops suddenly in the recombined material, and the effect is that of a recombination wave passing through the ejecta. This effect significantly alters the light curve evolution of Type II-P supernovae \citep[e.g.,][]{popov1993,Kasen:2009p17209}. 

During the passage of the recombination wave through the ejecta, the photosphere remains at the ionization temperature, $T_I$. The luminosity can then be calculated from the time-dependent photospheric radius: 

\begin{equation}
  L = 4\pi R_p^2 (t) \sigma T_I^4,
\end{equation}

\noindent where $R_p (t) = x_i (t) v t$ and $x_i$ is the dimensionless position of the photosphere in the expanding ejecta. We can write the equivalent of Equation (\ref{eq:2}) for the evolution of the internal energy of the ionized region, $\epsilon V_i = 4/3 \pi \epsilon (x_i v t)^3$:

\begin{equation} 
  \frac{1}{\epsilon}\frac{\partial \epsilon}{\partial t} + \frac{4}{x_i}\frac{\partial x_i}{\partial t} + \frac{4}{t} = \frac{H - L}{\epsilon V_i}.
\end{equation}

\noindent We again use the diffusion equation to write $L$ in terms of the internal energy, except now using $R_p$ instead of $R$. Finally, we equate the photospheric luminosity with that from diffusion in the ionized region, which gives an expression for $\epsilon$ in terms of $x_i$. The result is a non-linear first order differential equation for $x_i (t)$: 

\begin{equation}
\label{eq:4}
\frac{d x_i}{d t} = - \frac{2 x_i}{5 t} - \frac{t}{5 t_d^2 x_i} + \frac{1}{5 x_i^3 t} \left(\frac{H}{4\pi v^2 t_d^2 \sigma T_I^4}\right).
\end{equation}

\noindent In the absence of heating ($H=0$), Eq. (\ref{eq:4}) is similar to Eq. 14 of \citet{popov1993}, except with slightly different numerical coefficients. In this case, the analytic solution for the luminosity starting at time $t_i$, such that $x_i (t_i) = 1$, is:

\begin{equation}
L(t) = 4\pi\sigma T_I^4 v^2 \left[t_i^{6/5} t^{4/5} \left(1+\frac{t_i^2}{7 t_d^2}\right) - \frac{t^4}{7 t_d^2} \right].
\end{equation}

In general, we calculate the luminosity assuming constant opacity using Eq. (\ref{lcurve}). Then, the approximate one zone photospheric temperature is given by $\sigma T_{\rm p}^4 = L / 4\pi v^2 t^2$. When this drops below $T_I$, we numerically integrate Eq. (\ref{eq:4}) for $x_i (t)$, and then calculate $L(t) = 4 \pi v_f^2 t^2 x_i^2 \sigma T_I^4$. The recombination wave can significantly increase the peak luminosity in hydrogen rich progenitors (see Section \ref{outcomes}). More accurate radiative transfer calculations would likely find smoother light curves than those estimated from this one zone approach.

\end{document}